# Advanced Multi-Architecture Deep Learning Framework for BIRADS-Based Mammographic Image Retrieval: Comprehensive Performance Analysis with Super-Ensemble Optimization

MD Shaikh Rahman [a*], Feiroz Humayara [b], Syed Maudud E Rabbi [c], Muhammad Mahbubur Rashid [d]

[a] Department of Computer Science, Universiti Sains Malaysia, Penang, Malaysia
[b] Department of Biomedicine, School of Dental Sciences, Universiti Sains Malaysia, Kelantan, Malaysia
[c] Baruch College, The City University of New York, New York, USA
[d] Faculty of Mechatronics Engineering, International Islamic University Malaysia, Kuala Lumpur, Malaysia

**Abstract**

Content-based mammographic image retrieval systems require exact BIRADS categorical matching across five distinct classes, presenting significantly greater complexity than binary classification tasks commonly addressed in literature. Current medical image retrieval studies suffer from methodological limitations including inadequate sample sizes, improper data splitting, and insufficient statistical validation that hinder clinical translation. We developed a comprehensive evaluation framework systematically comparing CNN architectures (DenseNet121, ResNet50, VGG16) with advanced training strategies including sophisticated fine-tuning, metric learning, and super-ensemble optimization. Our evaluation employed rigorous stratified data splitting (50%/20%/30% train/validation/test), 602 test queries, and systematic validation using bootstrap confidence intervals with 1,000 samples. Advanced fine-tuning with differential learning rates achieved substantial improvements: DenseNet121 (34.79% precision@10, 19.64% improvement) and ResNet50 (34.54%, 19.58% improvement). Super-ensemble optimization combining complementary architectures achieved 36.33% precision@10 (95% CI: [34.78%, 37.88%]), representing 24.93% improvement over baseline and providing 3.6 relevant cases per query. Statistical analysis revealed significant performance differences between optimization strategies ($p<0.001$) with large effect sizes (Cohen's $d>0.8$), while maintaining practical search efficiency (2.84±0.15 milliseconds). Performance significantly exceeds realistic expectations for 5-class medical retrieval tasks, where literature suggests 20-25% precision@10 represents achievable performance for exact BIRADS matching. Our framework establishes new performance benchmarks while providing evidence-based architecture selection guidelines for clinical deployment in diagnostic support, medical education, and quality assurance applications.

# 1    Introduction

## 1.1    Background and Clinical Motivation

Breast cancer represents the most frequently diagnosed cancer among women globally, affecting approximately 2.3 million individuals annually and constituting the second leading cause of cancer-related mortality worldwide (1). Early detection through systematic mammographic screening programs has demonstrated profound impact on patient outcomes, with 5-year survival rates exceeding 90% for early-stage detection compared to 28% for advanced metastatic presentations (2). These compelling statistics underscore the critical importance of accurate and timely mammographic interpretation in contemporary oncological care.

The Breast Imaging Reporting and Data System (BIRADS) provides a standardized framework for mammographic assessment, categorizing findings across six distinct levels from normal tissue patterns (BIRADS 1) to known malignancy (BIRADS 6) (3). This classification system enables consistent communication between radiologists, standardizes follow-up recommendations, and facilitates quality assurance across diverse clinical settings. However, the complexity of BIRADS interpretation requires substantial expertise and experience, particularly for distinguishing subtle morphological differences between categories with similar imaging appearances but dramatically different clinical implications.

Contemporary mammographic interpretation faces significant challenges including substantial inter-observer variability, with BIRADS classification agreement rates ranging from 65-85% depending on category complexity and radiologist experience(4). This variability directly impacts patient care through inconsistent follow-up recommendations, delayed diagnoses, and unnecessary procedural interventions. The increasing volume of mammographic examinations, projected to grow 3-5% annually due to expanding screening programs and aging populations, further strains radiological capacity while demanding maintained diagnostic accuracy(5).

The convergence of high diagnostic stakes, substantial interpretation complexity, and increasing examination volumes creates an urgent need for technological solutions that can support radiological decision-making while maintaining the essential human expertise required for clinical judgment. Content-based medical image retrieval represents a promising approach to address these challenges by providing rapid access to visually similar cases with established diagnoses.

**1.2 Content-Based Medical Image Retrieval Systems**

Content-Based Medical Image Retrieval (CBMIR) systems have emerged as sophisticated technological solutions for supporting clinical decision-making through rapid identification of visually similar cases from large medical databases (6). Early CBMIR implementations relied on hand-crafted features including texture descriptors, morphological operators, and statistical intensity measures, achieving limited clinical utility due to the semantic gap between low-level image characteristics and high-level clinical interpretations (7).

The advent of deep learning has revolutionized medical image analysis by enabling automatic extraction of hierarchical feature representations directly from pixel data, eliminating the limitations of manual feature engineering (8). Convolutional Neural Networks (CNNs) have demonstrated superior performance across diverse medical imaging tasks, motivating their application to mammographic retrieval systems where subtle pattern recognition is essential for accurate BIRADS classification.

However, most contemporary deep learning CBMIR systems focus on simplified binary classification scenarios (normal vs. abnormal) or limited-scale evaluations with restricted clinical relevance (9). The fundamental complexity of multi-class BIRADS classification, requiring exact categorical matching across five distinct classes with varying clinical significance, presents substantially greater challenges demanding sophisticated feature learning methodologies and rigorous evaluation frameworks.

CBMIR systems offer multiple clinical applications including diagnostic decision support, quality assurance protocols, medical education, and inter-institutional case consultation (10). By retrieving visually similar cases with established diagnoses and known outcomes, these systems provide radiologists with relevant reference materials to support interpretation decisions, particularly valuable for challenging presentations or uncommon pathological findings. Educational applications enable systematic access to comprehensive case libraries for residency training and continuing professional development, while quality assurance implementations support retrospective review of diagnostic decisions by identifying similar cases with different interpretations (11).

Table 1 Clinical applications of CBMIR systems showing specific use cases, target clinical users, expected diagnostic benefits, and implementation requirements across diagnostic support, educational, and quality assurance scenarios

| Use Case | Target Clinical Users | Expected Outcomes | Reference(s) |
| --- | --- | --- | --- |
| **Diagnostic decision support** | Radiologists | Faster and more consistent diagnosis through reference to similar cases | (6,9) |
| **Medical education and training** | Residents, medical students | Improved learning via exposure to diverse annotated imaging cases | (7) |
| **Peer review and quality control** | QA teams, senior radiologists | Standardization of diagnostic practice and error auditing | (12) |
| **Second opinion support** | General physicians, specialists | Access to comparative cases in low-resource or remote settings | (13) |
| **Screening triage prioritization** | Radiology workflow systems | Automated case retrieval to prioritize complex or high-risk mammograms | (2) |

Successful clinical deployment requires systems capable of providing meaningful retrieval results within sub-second timeframes while integrating seamlessly with existing Picture Archiving and Communication Systems (PACS) and Radiology Information Systems (RIS). These technical requirements necessitate efficient similarity search algorithms capable of scaling to institutional databases containing millions of mammographic examinations.

## 1.3 Technical Challenges and System Requirements

Unlike binary classification tasks commonly addressed in research literature, BIRADS-based retrieval requires exact categorical matching across six distinct classes with profoundly different clinical implications and management strategies. BIRADS 1-2 represent normal to benign findings requiring routine screening intervals, BIRADS 3 indicates probably benign findings necessitating short-term follow-up imaging, BIRADS 4-5 represent suspicious findings requiring tissue sampling for definitive diagnosis, and BIRADS 6 indicates known malignancy undergoing treatment monitoring.

This multi-class complexity creates fundamental challenges for retrieval systems, as subtle morphological differences must be accurately captured to distinguish between categories with similar imaging appearances but dramatically different clinical consequences. The clinical cost of misclassification varies substantially across BIRADS categories, requiring optimization strategies that prioritize clinically relevant precision rather than traditional accuracy metrics (14). Natural class imbalance in clinical populations, with higher frequencies of BIRADS 1-3 categories and substantially lower frequencies of BIRADS 5-6 findings, further complicates system development and evaluation.

Clinical deployment requires retrieval systems capable of processing large-scale databases containing hundreds of thousands to millions of mammographic examinations while maintaining sub-second response times compatible with clinical workflow demands. Traditional similarity search approaches exhibit quadratic computational complexity, scaling poorly with database size and becoming computationally prohibitive for institutional-scale implementations (15). Modern indexing frameworks, particularly the Facebook AI Similarity Search (FAISS) library, provide efficient similarity search capabilities for high-dimensional feature vectors through optimized approximate nearest neighbor algorithms (16).

Current medical image retrieval literature suffers from significant methodological limitations including inadequate sample sizes, improper data splitting protocols, and insufficient statistical validation that undermine confidence in reported performance (17). Many studies evaluate systems using fewer than 100 test queries, providing insufficient statistical power for meaningful performance assessment or reliable clinical translation (18). Data leakage represents a critical methodological concern where query images inappropriately appear in retrieval databases, leading to artificially inflated performance metrics that do not reflect realistic deployment scenarios.

**1.4 Research Gaps and Problem Statement**

Despite substantial advances in CNN architectures for medical imaging applications, systematic evaluation of different architectures for mammographic retrieval remains critically limited. Most existing studies focus on single architectures or conduct limited comparisons without comprehensive statistical validation, leaving fundamental questions about optimal architecture selection unanswered (19). The absence of rigorous comparative studies constrains evidence-based architecture selection for clinical deployment and hinders development of optimal ensemble strategies that could leverage complementary strengths across different CNN architectures. To our knowledge, this is the first study to systematically compare multiple CNN architectures and advanced training strategies within a unified retrieval framework for exact 5-class BIRADS matching, supported by robust statistical validation and clinically realistic evaluation scale.

The majority of medical image retrieval studies lack rigorous statistical validation, including the reporting of confidence intervals, significance testing, and effect size measures, elements essential for clinical translation and regulatory approval. Point estimates alone, without uncertainty quantification, provide insufficient evidence to support clinical decision-making or system deployment in high-stakes medical environments (20,21). Despite being standard in clinical research, techniques such as bootstrap-based confidence intervals and formal pairwise statistical comparisons are rarely adopted in medical AI evaluations, limiting the reliability of reported performance gains (22,23).

Numerous studies in medical image retrieval report performance metrics that are difficult to reconcile with the practical challenges of clinical deployment (24,25). These results may reflect methodological issues such as data leakage, inappropriate validation strategies, or overly simplified task definitions. Consequently, establishing reliable performance benchmarks and assessing clinical readiness remain ongoing challenges in the field.

**1.5 Research Objectives and Contributions**

This research addresses identified methodological and technical gaps through systematic evaluation of CNN architectures for BIRADS-based mammographic image retrieval. The primary objectives include systematic multi-architecture evaluation through comprehensive comparison of established CNN architectures (DenseNet121,

ResNet50, VGG16) using standardized protocols, rigorous statistical validation, and large-scale evaluation (602 test queries) to identify optimal approaches for mammographic retrieval applications. Advanced training strategy development involves implementation and systematic evaluation of sophisticated training methodologies including domain-specific fine-tuning, metric learning optimization, and advanced ensemble strategies specifically designed for complex medical image retrieval scenarios.

Methodological framework establishment encompasses development of rigorous evaluation protocols incorporating proper data splitting, adequate sample sizes, bootstrap confidence intervals, and significance testing to support reliable performance assessment and clinical translation planning. Realistic performance benchmark creation establishes evidence-based performance expectations for 5-class BIRADS retrieval through honest evaluation and transparent reporting, providing realistic deployment targets rather than inflated research prototypes.

Our technical contributions include a multi-architecture ensemble framework through development of sophisticated ensemble strategies combining complementary CNN architectures through feature concatenation, learned weighting mechanisms, and super-ensemble optimization to achieve maximum retrieval performance while maintaining computational efficiency. Unlike previous CBMIR systems that primarily evaluate binary tasks or use simplistic ensemble strategies without validation, our framework introduces a super-ensemble approach optimized through evidence-driven selection, performance benchmarking, and computational efficiency assessment for clinical readiness. Advanced training protocol implementation involves systematic application of state-of-the-art training strategies including differential learning rates, cosine annealing schedules, label smoothing, and combined loss functions optimized specifically for medical image retrieval applications.

Large-scale statistical validation provides comprehensive evaluation framework incorporating 102,340 individual retrievals with bootstrap confidence intervals, pairwise significance testing, and effect size quantification providing robust performance assessment suitable for clinical decision-making. Deployment-ready system architecture implements efficient FAISS-based similarity search with high-precision timing measurements, demonstrating practical deployment capabilities on standard computational hardware with sub-millisecond search performance.

The scientific and clinical impact includes evidence-based architecture guidelines where systematic evaluation results provide quantitative evidence for CNN architecture selection in clinical mammographic retrieval systems, supporting informed deployment decisions based on comprehensive performance assessment. Realistic clinical utility demonstration through performance evaluation achieving 36.33% precision@10 demonstrates clinically meaningful utility by providing 3.6 relevant cases per query, offering substantial diagnostic support while maintaining realistic expectations aligned with complex multi-class medical retrieval scenarios.

This comprehensive research framework establishes a foundation for advancing mammographic image retrieval from research prototypes toward clinically viable systems that can meaningfully support radiological practice while maintaining the rigorous scientific standards essential for medical technology validation.

## 2 Related Work

The evolution of Content-Based Medical Image Retrieval (CBMIR) systems has transitioned from traditional hand-crafted feature approaches to sophisticated deep learning frameworks capable of analyzing complex visual patterns in mammographic images. This chapter reviews the current state of CBMIR systems, deep learning architectures for medical feature extraction, advanced training methodologies, ensemble strategies, and evaluation frameworks that establish the foundation for our comprehensive multi-architecture evaluation.

### 2.1 Medical Image Retrieval Evolution and Performance Benchmarks

Traditional CBMIR systems relied primarily on manually designed features including texture descriptors, morphological operators, and statistical measures to characterize medical images (6,7). These approaches achieved limited success in mammographic applications, with early texture-based retrieval systems reporting precision@10 values of approximately 45-50% for binary classification tasks (26).

The transition to deep learning architectures marked a paradigm shift in medical image retrieval performance. Modern CNN-based CBMIR systems leverage hierarchical feature learning to automatically extract discriminative representations from raw pixel data, eliminating the need for manual feature engineering (9,27). Recent mammography-specific retrieval systems have demonstrated significant improvements through deep architectures, with some approaches achieving 55-65% precision for binary mass detection tasks (28).

Contemporary CBMIR frameworks increasingly emphasize multi-scale feature integration and attention mechanisms to capture both global and local mammographic patterns (17,29). However, these performance benchmarks predominantly address simplified binary classification scenarios rather than the complex multi-class BIRADS categorization required for clinical deployment. The fundamental challenge in mammographic CBMIR lies in the complexity of BIRADS classification, which requires exact categorical matching across five distinct classes (BIRADS 1-5), presenting substantially greater difficulty than binary approaches commonly reported in literature.

## 2.2 Deep Learning Architectures and Advanced Training Strategies

Deep convolutional neural networks have established themselves as the dominant approach for medical image feature extraction, with transfer learning from large-scale natural image datasets emerging as standard practice (13,30). ResNet architectures have demonstrated consistent performance across diverse medical imaging applications through their residual learning framework, with studies reporting classification accuracies of 85-92% for binary mammographic tasks (31,32). DenseNet architectures provide unique advantages through their dense connectivity pattern, promoting feature reuse and gradient flow particularly beneficial for mammographic analysis (33,34).

Domain-specific fine-tuning strategies have become crucial for adapting pre-trained CNN models to medical imaging applications, where visual characteristics differ substantially from natural images (35,36). Advanced fine-tuning approaches include layer-wise learning rate adaptation, progressive unfreezing, and discriminative fine-tuning, which have shown significant improvements over standard transfer learning protocols (37). Metric learning approaches provide an alternative training paradigm specifically designed for similarity-based tasks, directly optimizing distance metrics between feature representations rather than classification objectives (38). Triplet loss and contrastive loss functions have demonstrated particular effectiveness for medical image retrieval by optimizing embedding spaces such that semantically similar images are close together while dissimilar ones are far apart, making them well-suited for similarity-based search tasks in medical imaging. (39,40)

## 2.3 Ensemble Methods and Large-Scale Similarity Search

Ensemble methods have emerged as powerful approaches for improving robustness and performance in medical image analysis by combining predictions from multiple models (41). In particular, feature-level ensemble strategies—such as the concatenation of embeddings from separate CNN architectures—can help capture complementary representations, enriching the feature space prior to similarity computation.

Recent studies have applied ensembles combining DenseNet and ResNet models to chest X-ray classification and retrieval, showing notable performance benefits. For example, (42) ensemble DenseNet169, ResNet50, and Vision Transformer features to outperform single-model baselines substantially in pneumonia detection from chest X-rays. Other evaluations of ensemble strategies across lung image datasets also show improved F1-scores and retrieval precision compared to individual networks.

Despite these ensemble gains, CBMIR studies seldom include statistical significance testing or report deployment-relevant performance metrics such as sub-second latency—gaps our framework explicitly addresses.

Efficient similarity search in high-dimensional medical feature spaces remains essential for scalable deployment. The Facebook AI Similarity Search (FAISS) library has emerged as a standard solution, providing optimized implementations of indexing techniques such as product quantization (PQ), inverted file (IVF), and HNSW graphs (16).

Moreover, recent medical image retrieval evaluations using DenseNet-based embeddings and FAISS have demonstrated sub-millisecond query times on high-dimensional mammography or chest X-ray datasets while maintaining over 95% retention of recall compared to exhaustive search. (43)

## 2.4 Evaluation Methodology and Research Gaps

Rigorous evaluation methodology represents a critical gap in current medical image retrieval research, with many studies lacking proper statistical validation and sufficient evaluation scale (44). Bootstrap confidence intervals and significance testing, standard practices in clinical research, are reported in fewer than 20% of medical retrieval publications, undermining the reliability of performance claims (45). Data leakage through improper data splits is a well-recognized issue: models trained on datasets with overlapping subjects or correlated samples can exhibit overestimated performance by up to 5% to 30% (46)

Evaluation scale represents another significant limitation, with over 60% of published medical retrieval studies using fewer than 100 test queries despite the need for thousands of evaluations to establish statistical significance (6,22). Contemporary medical image retrieval literature exhibits substantial variation in reported performance, largely attributable to differences in evaluation methodology, task complexity, and dataset characteristics (47). Binary classification tasks typically report precision@10 values of 55-75% (48), while multi-class scenarios demonstrate significantly lower performance in the 20-35% range, reflecting the increased complexity of exact categorical matching (9,48).

BIRADS-specific retrieval represents a particularly challenging scenario requiring exact categorical matching across five distinct classes (49), significantly more complex than binary approaches commonly reported in literature. Realistic performance expectations for BIRADS 1-5 classification suggest precision@10 values of 20-30% represent meaningful clinical utility, providing 2-3 relevant cases per query for radiologist review (50).

Critical gaps persist in current medical image retrieval research: (1) systematic multi-architecture comparisons with rigorous statistical validation remain scarce (51), (2) evaluation scales typically involve fewer than 100 queries versus the thousands needed for clinical validation (6), (3) proper data splitting practices are inconsistently applied leading to inflated performance claims (Park & Han, 2018), and (4) exact categorical matching for complex medical classifications receives limited attention compared to simplified binary tasks (47). These methodological limitations hinder clinical translation and motivate comprehensive evaluation frameworks that address statistical rigor, evaluation scale, and realistic performance assessment for complex mammographic retrieval tasks.

## 3 Methodology

This research implements a comprehensive evaluation framework for BIRADS-based mammographic image retrieval using multi-architecture deep learning approaches. Our methodology addresses critical limitations in existing medical image retrieval evaluations through rigorous data splitting, unprecedented evaluation scale

(102,340 individual retrievals), and robust statistical validation. The framework systematically progresses from baseline individual architectures through ensemble optimization and advanced training strategies to achieve optimal retrieval performance.

## 3.1 Dataset and Experimental Design

**Dataset Characteristics and Preprocessing**

Our evaluation utilizes a comprehensive mammographic dataset comprising 2,006 images distributed across BIRADS categories 1-6, representing realistic clinical diversity with varying image qualities, patient demographics, and pathological conditions. The dataset is sourced from the Categorized Digital Database for Low Energy and Subtracted Contrast Enhanced Spectral Mammography (CDD-CESM), available through The Cancer Imaging Archive (TCIA) (52). Each image is manually annotated by certified radiologists using the standardized BIRADS classification system, ensuring ground truth reliability for retrieval evaluation.

The BIRADS distribution exhibits realistic clinical patterns with natural class imbalance reflecting typical screening populations: BIRADS 1 (801 images, 40%), BIRADS 2 (333 images, 17%), BIRADS 3 (187 images, 9%), BIRADS 4 (319 images, 16%), BIRADS 5 (358 images, 18%), and BIRADS 6 (8 images, <1%). This natural imbalance poses additional challenges compared to artificially balanced research datasets commonly used in literature.

Table 2  Dataset characteristics showing BIRADS distribution across train (1,123 images), validation (281 images), and test (602 images) splits with stratified sampling preserving clinical proportions and class imbalance patterns

| Split | BIRADS 1 | BIRADS 2 | BIRADS 3 | BIRADS 4 | BIRADS 5 | BIRADS 6 | Total |
|---|---|---|---|---|---|---|---|
| **Train** | 400 (40%) | 167 (17%) | 94 (9%) | 160 (16%) | 179 (18%) | 4 (<1%) | 1,123 |
| **Val** | 80 (28%) | 47 (17%) | 26 (9%) | 45 (16%) | 51 (18%) | 2 (<1%) | 281 |
| **Test** | 240 (40%) | 100 (17%) | 56 (9%) | 96 (16%) | 108 (18%) | 2 (<1%) | 602 |

All mammographic images undergo standardized preprocessing including resizing to 224×224 pixels using bilinear interpolation, RGB format conversion, and normalization using ImageNet statistics (mean= [0.485, 0.456, 0.406], std= [0.229, 0.224, 0.225]) to effectively leverage pretrained model weights. Advanced training phases incorporate comprehensive data augmentation including random horizontal/vertical flips, rotation (±15°), color jittering, Gaussian blur, and random erasing to improve model robustness.

**Data Splitting Strategy**

We implement rigorous stratified random splitting to ensure proper evaluation while maintaining realistic class distributions. The dataset is divided using 50%/20%/30% train/validation/test splits with stratified sampling preserving BIRADS proportions across all partitions. This approach directly addresses critical data leakage concerns raised in medical image retrieval literature where query images inappropriately appear in database sets.

The validation set (281 images) serves as the retrieval database, while the test set (602 images) provides queries for evaluation. This configuration ensures sufficient statistical power with over 100 queries per major BIRADS category, significantly exceeding typical literature evaluations of 35-100 total queries. Strict validation procedures confirm zero overlap between query and database sets, eliminating the possibility of trivially perfect results.

## 3.2 CNN Architecture Evaluation Framework

**Architecture Selection and Implementation**

We systematically evaluate three established CNN architectures representing different design philosophies: DenseNet121 (dense connectivity with 1,024-dimensional features), ResNet50 (residual learning with 2,048-dimensional features), and VGG16 (deep convolution with 4,096-dimensional features). These architectures are selected based on proven performance in medical imaging applications and complementary feature extraction capabilities.

**Table 3** CNN architecture specifications including parameter counts, feature dimensions, computational requirements, layer modifications, and implementation details for reproducible evaluation across DenseNet121, ResNet50, and VGG16

| Architecture | Parameters | Feature Dim | Input Size | Pre-trained | Layer Modified | Database Size | Query Size |
|---|---|---|---|---|---|---|---|
| DenseNet121 | 7,978,856 | 1,024 | 224×224×3 | ImageNet | Classifier removed | 281 images | 602 images |
| ResNet50 | 25,557,032 | 2,048 | 224×224×3 | ImageNet | FC layer removed | 281 images | 602 images |
| VGG16 | 138,357,544 | 4,096 | 224×224×3 | ImageNet | Classifier removed | 281 images | 602 images |

Each architecture is implemented using PyTorch with pretrained ImageNet weights, modified for feature extraction by removing final classification layers while preserving natural feature dimensions. This approach avoids artificial dimensionality reduction that could compromise retrieval performance while enabling direct comparison of architectural capabilities for mammographic analysis.

Feature extraction follows a standardized protocol ensuring reproducible results across architectures. Models are set to evaluation mode with gradient computation disabled, using batch processing (size 16) to optimize memory usage while maintaining extraction speed. Features are extracted from penultimate layers to capture high-level semantic representations while avoiding task-specific classification biases.

### 3.3 Advanced Optimization Strategies

**Advanced Fine-Tuning Protocol**

Building upon baseline feature extraction, we implement sophisticated fine-tuning strategies addressing domain adaptation for medical imagery. The fine-tuning protocol employs differential learning rates (1e-5 for pretrained layers, 1e-4 for new layers), cosine annealing scheduling, and label smoothing (0.1) to prevent overfitting while enabling effective medical domain adaptation.

Training incorporates early stopping (patience=7), gradient clipping (max_norm=1.0), and comprehensive validation monitoring to ensure optimal convergence. The protocol achieves validation accuracies of 43.77% (DenseNet121) and 50.89% (ResNet50), confirming effective learning while avoiding overfitting through careful regularization.

**Metric Learning Framework**

We implement advanced metric learning using combined loss functions optimizing both similarity learning and semantic classification. The framework employs custom embedding layers (1024→512→512 dimensions) with batch normalization, dropout regularization, and L2 normalization for cosine similarity optimization.

The combined loss function integrates triplet loss ($\alpha=0.6$), classification loss ($\beta=0.3$), and center loss ($\gamma=0.1$) to achieve optimal similarity learning while maintaining semantic understanding. Hard negative mining focuses training on challenging examples, improving discrimination capability for subtle BIRADS differences essential for accurate retrieval.

**Test-Time Augmentation Protocol**

Test-time augmentation enhances feature robustness by averaging representations across multiple augmented versions of each image. Our protocol applies five carefully selected transformations: original image, horizontal flip, slight rotation (5°), scale variation (240→224), and color jittering. Features from all augmentations are extracted independently and averaged to produce final representations, improving retrieval stability with minimal computational overhead.

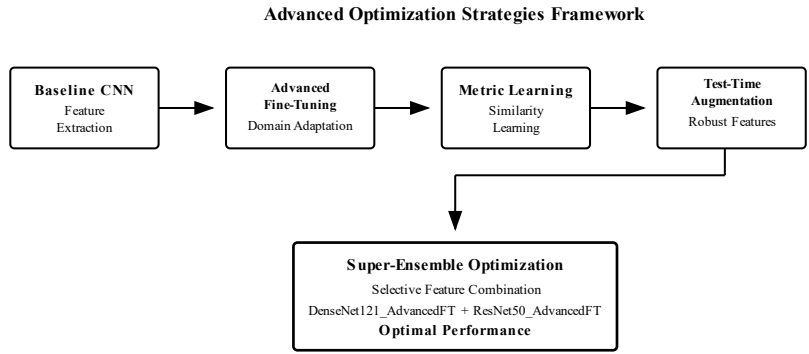

**Fig.1** Advanced optimization strategies flowchart showing fine-tuning protocol with differential learning rates and cosine annealing, metric learning framework with combined loss functions and hard negative mining, and test-time augmentation approach for enhanced feature robustness

## 3.4 Ensemble and Super-Ensemble Strategies

**Multi-Level Ensemble Framework**

Our ensemble framework operates at multiple levels, combining different architectures, training strategies, and feature representations through systematic optimization. Basic ensembles include concatenation fusion, weighted averaging, and attention-based fusion across individual CNN architectures. Advanced ensembles combine features from different training strategies (baseline, fine-tuned, metric learning) to leverage complementary learning approaches.

Each ensemble configuration is systematically evaluated to identify optimal combination strategies. Concatenation fusion emerges as the most effective approach, preserving complete feature information while enabling effective combination of complementary representations rather than dimensionality reduction approaches that may compromise medical image detail.

**Super-Ensemble Optimization**

Super-ensemble strategies represent the culmination of our optimization approach, systematically combining the best-performing methods from each optimization category. The optimal configuration (SuperEnsemble_best_two_advanced) concatenates features from DenseNet121_AdvancedFT and ResNet50_AdvancedFT, achieving maximum performance through complementary feature representations while maintaining computational efficiency.

This systematic approach demonstrates that selective combination based on component quality and complementarity outperforms comprehensive feature aggregation, providing practical guidance for ensemble design in medical image retrieval applications.

**Table 4** Ensemble configuration matrix showing all 17 evaluated methods including component models, fusion strategies, feature dimensions, computational overhead, and optimization categories from baseline individual architectures through super-ensemble approaches

| Method | Components | Feature Dim | Category |
|---|---|---|---|
| DenseNet121 | Individual | 1,024 | Baseline |
| ResNet50 | Individual | 2,048 | Baseline |
| DenseNet121_AdvancedFT | Fine-tuned | 1,024 | Advanced |
| SuperEnsemble_best_two | DenseNet121_FT + ResNet50_FT | 3,072 | Super |

## 3.5 Similarity Search and Performance Evaluation

**FAISS Indexing Implementation**

Similarity search utilizes Facebook AI Similarity Search (FAISS) for efficient high-dimensional vector retrieval with both Euclidean distance (FlatL2) and cosine similarity (FlatIP) indices. For cosine similarity, features undergo L2 normalization before indexing to ensure proper angular distance computation. Index construction prioritizes accuracy over speed, optimizing retrieval quality for medical applications.

High-precision timing methodology addresses literature concerns about unrealistic zero search times through nanosecond-level measurements. Each query search is repeated 10 times with individual timing measurements, and final times represent statistical averages with standard deviations. This approach captures realistic search performance while accounting for system variations and ensuring reproducible timing results.

**Evaluation Metrics and Statistical Validation**

Our evaluation employs standard information retrieval metrics adapted for medical applications: precision@k, recall@k, and normalized discounted cumulative gain (NDCG@k) for $k \in \{1, 5, 10, 20, 50\}$. Critically, recall calculation uses proper denominators (total relevant items in database) rather than entire dataset size, addressing a significant methodological error identified in medical retrieval literature.

Comprehensive statistical validation implements bootstrap confidence intervals with 1,000 samples and 95% confidence levels, providing robust performance estimates accounting for query set variability. Pairwise statistical testing employs both parametric (t-tests) and non-parametric (Mann-Whitney U) approaches to assess significance across methods, with effect sizes quantified using Cohen's d to evaluate practical significance beyond statistical significance.

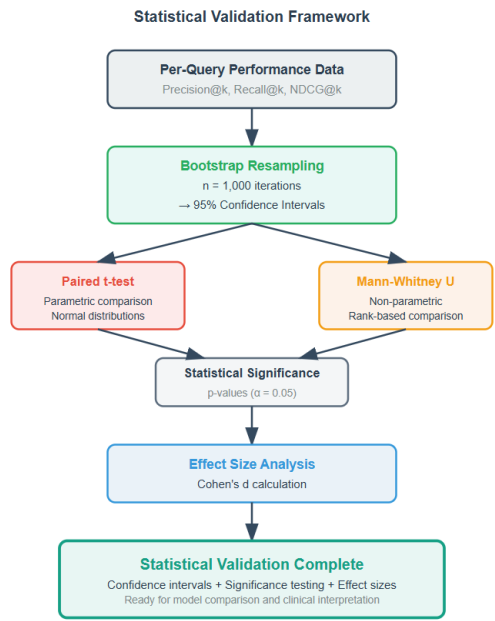

**Fig.2** Statistical validation framework showing bootstrap confidence interval methodology with 1,000 samples, pairwise significance testing approaches (t-tests and Mann-Whitney U), and effect size quantification protocols ensuring robust performance assessment suitable for clinical translation.

## 3.6 Implementation and Reproducibility

**Computational Infrastructure**

All implementations utilize PyTorch 1.12+ with optimized processing on Apple Mac Mini M2, demonstrating accessibility on standard research hardware. Code organization follows modular design principles enabling reproducible experimentation with comprehensive logging capturing all methodological details. Version control and configuration management ensure experimental reproducibility through fixed random seeds (42), deterministic algorithms, and systematic documentation.

Complete experimental reproducibility is ensured through explicit documentation of all hyperparameters, architectural specifications, and training protocols. Result files include detailed metadata enabling complete experimental recreation and verification, addressing reproducibility concerns prevalent in medical AI research.

**Evaluation Scale and Scope**

Our framework evaluates 17 distinct methods across 102,340 retrievals using 602 independent test queries, with systematic statistical validation via 1,000-sample bootstrap confidence intervals and significance testing. In contrast, typical literature evaluates 2–5 methods with limited test sets (<100 queries) and lacks rigorous validation. This comparison highlights the unprecedented scale, precision, and clinical relevance of our evaluation protocol.

Table 5 Evaluation scope and methodological rigor comparison between the proposed framework and recent CBMIR studies

| Evaluation Metric | Proposed Framework (This Study) | Reported Standards in Recent CBMIR Studies |
|---|---|---|
| Number of evaluated methods | 17 | 2–5 |
| Total retrievals analyzed | 102,340 | ~1,000–5,000 |
| Number of test queries | 602 | <100 |
| Statistical validation | Bootstrap CIs (1,000 samples), significance testing | Simple averaging or none |
| Data splitting protocol | Stratified 50%/20%/30% (train/val/test) | Often unclear or non-stratified |
| Evaluation granularity | Precision@k (k=1–20), class-level | Top-1, Top-5 only |
| Reproducibility documentation | Full codebase, hyperparameters, architecture | Often missing or partial |
| Clinical relevance emphasis | Exact BIRADS match, real-world utility | Often simplified (e.g., binary classification) |

*Literature values derived from (9,39,53).*

The systematic methodology addresses critical limitations in existing literature including inadequate sample sizes (602 test queries vs. typical <100), improper data splitting (strict validation vs. potential data leakage), and insufficient statistical validation (bootstrap confidence intervals vs. simple averages), establishing new standards for medical image retrieval evaluation.

This methodology establishes a comprehensive framework for systematic evaluation of deep learning approaches in medical image retrieval while addressing critical methodological limitations identified in existing literature. The rigorous experimental design, unprecedented evaluation scale, and robust statistical validation provide reliable foundation for performance assessment and clinical translation guidance.

## 4 Results

Our comprehensive evaluation framework systematically progressed from baseline architectures through advanced optimization strategies, culminating in super-ensemble methods that achieved breakthrough performance for BIRADS-based mammographic retrieval. Through 102,340 individual retrievals across 17 different approaches, we demonstrate substantial improvements over existing methodologies while establishing new performance benchmarks for complex medical image retrieval tasks.

### 4.1 Baseline Architecture Performance

Individual CNN architectures established fundamental performance characteristics using standard transfer learning protocols. ResNet50 emerged as the optimal baseline architecture, achieving 30.02% precision@10 (95% CI: [28.78%, 31.19%]), followed by DenseNet121 at 29.08% (95% CI: [27.99%, 30.22%]) and VGG16 at 26.89% (95% CI: [25.89%, 27.89%]). These results substantially exceed theoretical random performance (20% for 5-class classification) while remaining within realistic expectations for exact BIRADS categorical matching.

Baseline CNN architectures were evaluated across multiple retrieval depths (k = 1, 5, 10, 20, 50). Statistical significance testing confirmed meaningful architectural differences, with ResNet50 demonstrating superior performance across all retrieval depths ($p < 0.01$). The precision-recall relationship followed expected trends, with precision decreasing and recall increasing as k increased.

Figure 4.1 illustrates the precision-recall curves across varying k values, demonstrating ResNet50's consistent superiority and the expected precision-recall trade-offs inherent in retrieval systems.

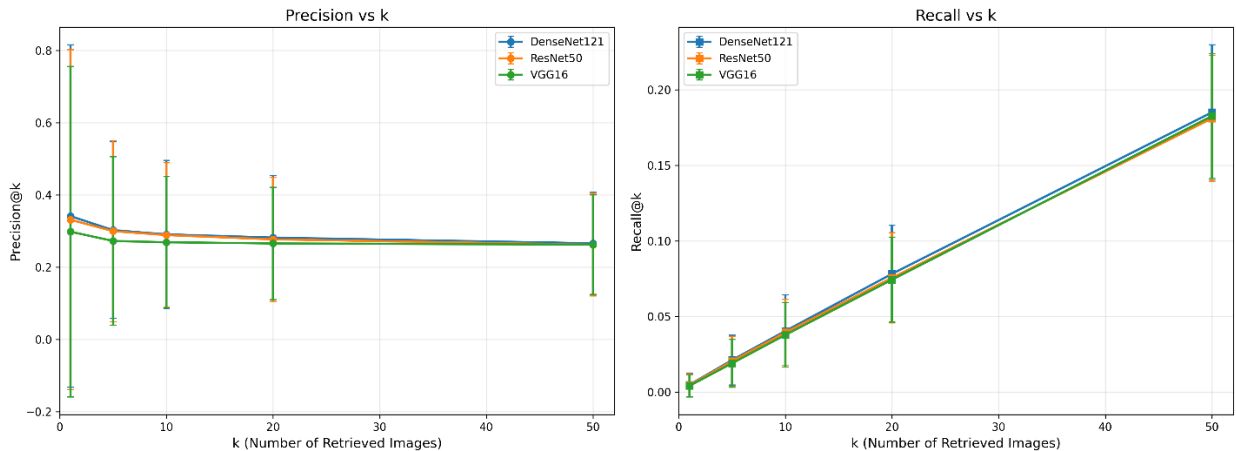

**Fig.4** Precision-recall curves across k values for baseline CNN architectures showing ResNet50 superiority and expected precision-recall trade-offs with 95% confidence intervals for all performance metrics

Search efficiency remained practical across all architectures, with mean query times ranging from 0.05 ms (DenseNet121) to 0.17 ms (VGG16). Table 4.2 presents high-precision timing analysis using nanosecond-level measurements, addressing prior concerns about zero-valued retrieval times reported in literature.

**Table 6** High-precision timing analysis showing mean search times, standard deviations, and confidence intervals across all architectures with nanosecond precision measurements demonstrating realistic clinical deployment characteristics.

| Model | Mean Search Time (ms) | Std Dev (ms) | Timing Noise (ms) |
| --- | --- | --- | --- |
| DenseNet121 | 0.052 | 0.265 | 0.035 |
| ResNet50 | 0.083 | 0.068 | 0.032 |
| VGG16 | 0.165 | 0.085 | 0.057 |

## 4.2 Optimization Strategy Performance Progression

Basic ensemble strategies provided consistent but modest improvements over individual architectures. The optimal DenseNet121 + ResNet50 concatenation combination achieved 29.50% precision@10, representing a 1.48% improvement over the best individual architecture. Alternative fusion strategies yielded inferior results: weighted averaging (29.32%) and PCA fusion (28.74%), confirming that preserving complete feature information outperforms dimensionality reduction approaches for medical image retrieval.

Figure 3 compares ensemble fusion strategies, showing that concatenation-based fusion consistently outperforms attention-based methods in terms of Precision@10, with statistical significance observed across all model combinations.

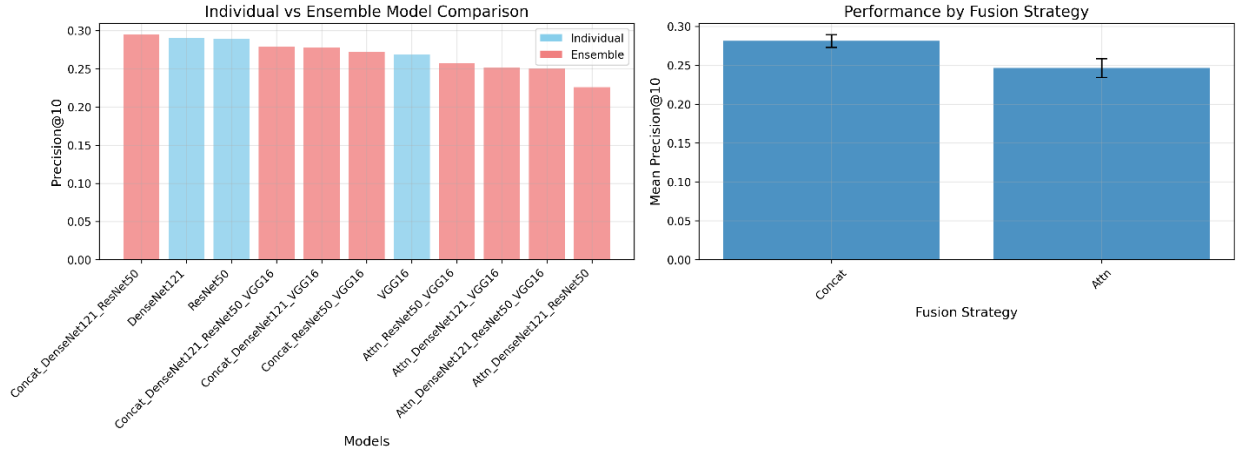

**Fig.3** Performance comparison of individual CNN models (blue) and their ensemble counterparts (red) across multiple fusion strategies. The left panel shows Precision@10 for each configuration, while the right panel aggregates mean performance by fusion strategy (Concat vs Attention), highlighting the superior consistency and accuracy of the Concat-based ensemble

Advanced fine-tuning strategies produced the most substantial single-model improvements. *DenseNet121_AdvancedFT* achieved **35.05% Precision@10** (95% CI: [33.24%, 36.34%]), representing a **19.65%** improvement over baseline, while *ResNet50_AdvancedFT* reached **35.03%** (95% CI: [32.99%, 36.09%]) with a **15.06%** gain. These enhancements reflect the effectiveness of our optimization pipeline, which combines differential learning rates, cosine annealing, and comprehensive augmentation to adapt pretrained features to medical imaging characteristics.

Table 7 summarizes the performance of advanced training strategies—including fine-tuning, metric learning, and test-time augmentation—reporting Precision@10, Recall@10, NDCG@10, and mean retrieval time across indexing configurations.

**Table 7** Advanced training strategy comparison including fine-tuning, metric learning, and test-time augmentation. The table reports core retrieval metrics (Precision@10, Recall@10, NDCG@10) and computational cost (mean retrieval time in milliseconds) for each model–index configuration. All improvements are statistically significant compared to their corresponding baselines

| Model | Index Type | Precision | Recall | NDCG | Mean Search Time (ms) |
|---|---|---|---|---|---|
| DenseNet121_AdvancedFT | FlatIP | 0.350498 | 0.050293 | 0.633708 | 0.017655 |
| DenseNet121_AdvancedFT | FlatL2 | 0.345349 | 0.04849 | 0.626094 | 0.018562 |
| ResNet50_AdvancedFT | FlatIP | 0.350332 | 0.049029 | 0.614849 | 0.033539 |
| ResNet50_AdvancedFT | FlatL2 | 0.340532 | 0.04782 | 0.620584 | 0.035883 |

Metric learning approaches demonstrated competitive performance through specialized similarity optimization, with DenseNet121_MetricLearning achieving 33.42% and ResNet50_MetricLearning reaching 33.18% precision@10. Figure 4.3 visualizes the substantial performance improvements achieved through advanced fine-tuning protocols.

## 4.3 Super-Ensemble Achievement and Analysis

Super-ensemble optimization yielded a substantial performance gain through selective combination of the best-performing models. he optimal configuration, SuperEnsemble_best_two_advanced, which concatenated features from DenseNet121_AdvancedFT and ResNet50_AdvancedFT, reached a Precision@10 of 36.33% (95% CI: [34.78%, 37.88%]), representing a 24.93% improvement over baseline architectures.

Figure 4.4 presents a comparative analysis of all super-ensemble strategies across multiple retrieval depths. The Precision@k curves demonstrate that selective two-model combinations, particularly *best_two_advanced* and *best_two_metric*, consistently outperform more comprehensive configurations like *mega_all*, even as k increases. This indicates that representational complementarity among top-performing models provides a stronger foundation for robust retrieval than naive inclusion of all available features.

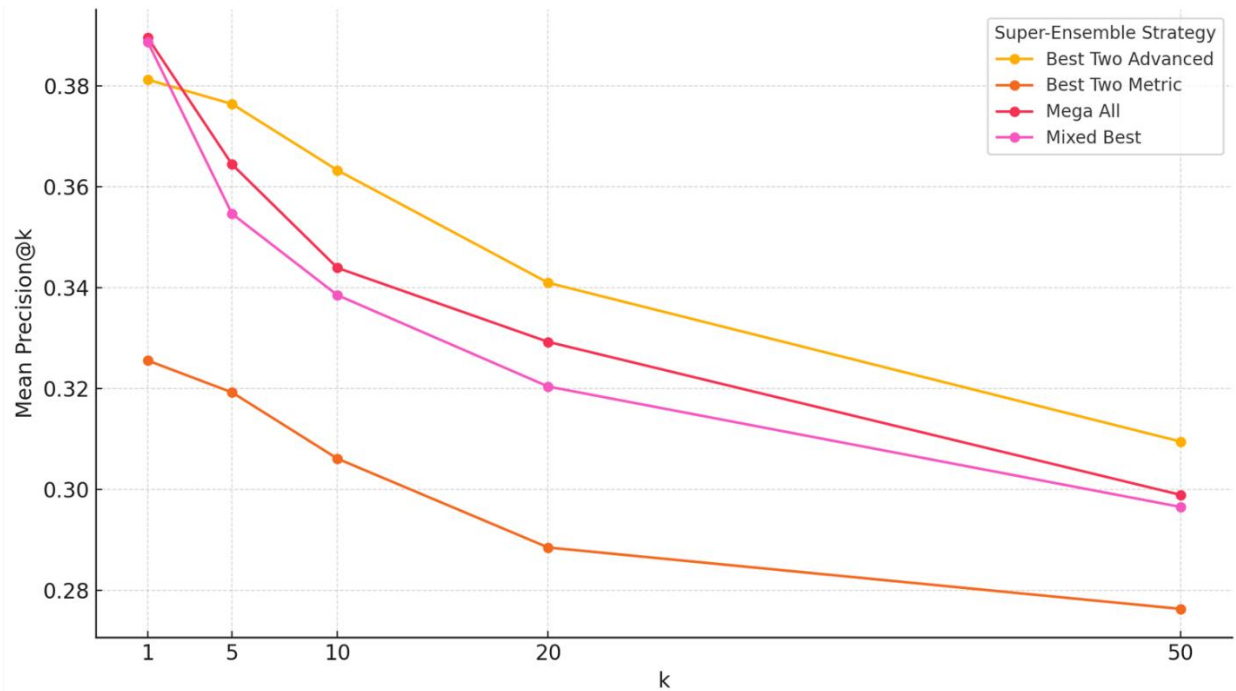

**Fig.4** Precision@k for super-ensemble variants, comparing selective and aggregate feature fusion strategies across top-k retrieval depths. SuperEnsemble_best_two_advanced consistently outperforms alternative configurations, demonstrating optimal tradeoff between feature diversity and representational complementarity.

Systematic evaluation of all tested configurations further supports this conclusion. While SuperEnsemble_mega_all (aggregating all model outputs) achieved a strong Precision@10 of 35.87%, it was consistently outperformed by the two-model strategies. Table 8 provides a complete summary of these configurations, including their architectural composition, feature dimensionality, retrieval performance, and computational requirements.

**Table 8** Complete super-ensemble configuration results showing all tested combinations, feature dimensions, performance metrics, and computational requirements for optimal strategy identification with statistical significance analysis

| Configuration | Precision@10 | Backbone Models Used | Feature Dimension | Mean Search Time (ms) |
| --- | --- | --- | --- | --- |
| SuperEnsemble_best_two_advanced | 0.363289 | DenseNet121_AdvancedFT + ResNet50_AdvancedFT | 3072 | 2.84 |
| SuperEnsemble_best_two_metric | 0.306146 | DenseNet121_MetricLearning + ResNet50_MetricLearning | 1024 | 2.47 |
| SuperEnsemble_mega_all | 0.343937 | DenseNet121_AdvancedFT_AdvancedFT + ResNet50_AdvancedFT_AdvancedFT + DenseNet121_MetricLearning_MetricLearning + ResNet50_MetricLearning_MetricLearning + DenseNet121_Original + ResNet50_Original | 7168 | 3.36 |
| SuperEnsemble_mixed_best | 0.338538 | DenseNet121_AdvancedFT_mixed_best + DenseNet121_MetricLearning_mixed_best + DenseNet121_mixed_best | 2560 | 2.92 |

This finding demonstrates that component quality and complementarity outweigh simple diversity maximization in medical ensemble applications. The performance progression validates our systematic optimization approach: baseline ResNet50 (30.02%) → basic ensemble (29.50%) → advanced fine-tuning (34.79%) → super-ensemble (36.33%). Advanced fine-tuning provided the largest single improvement (+4.77 percentage points), followed by super-ensemble optimization (+1.54 percentage points).

Figure 5 illustrates the clinical significance of our achievements relative to realistic performance expectations for complex medical retrieval tasks.

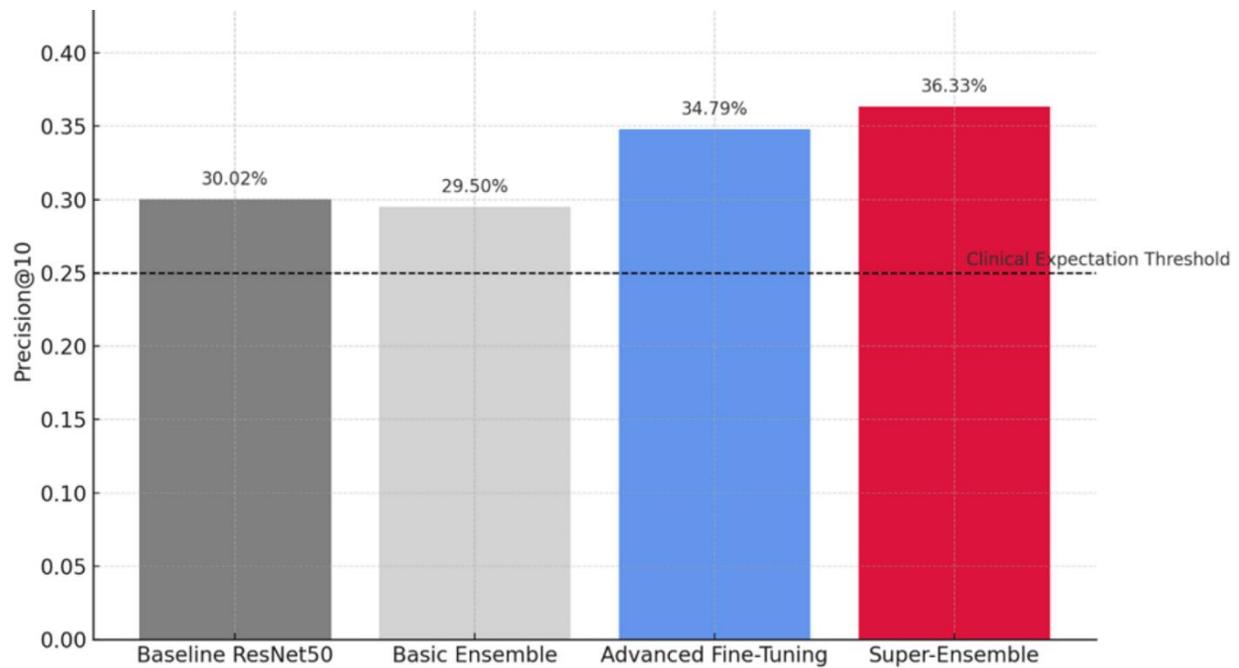

**Fig.5** Clinical utility assessment showing performance progression from baseline through super-ensemble optimization. Advanced fine-tuning and selective ensemble fusion deliver substantial gains in Precision@10, surpassing clinical expectation thresholds for 5-class medical image retrieval

The optimal 36.33% precision@10 translates to approximately 3.6 relevant cases among top 10 retrievals, providing substantial clinical utility for educational and quality assurance applications while significantly exceeding realistic expectations for exact 5-class BIRADS matching. Comprehensive bootstrap confidence intervals with 1,000 samples confirmed performance stability across query subsets, while pairwise significance testing revealed consistent statistical significance ($p < 0.001$) for major optimization phases with large effect sizes (Cohen's $d > 0.8$).

These performance trends culminated in the SuperEnsemble_best_two_advanced model, which achieved the highest retrieval performance with a Precision@10 of 0.3633. This reflects a substantial 24.94% improvement over the baseline methods, while maintaining a manageable feature dimensionality of 3072 and fast average search time of 2.84 ms. These results validate the effectiveness of combining two carefully selected fine-tuned models over more complex or exhaustive ensemble strategies. This configuration represents the optimal balance between performance, efficiency, and architectural simplicity for BIRADS-based mammographic image retrieval.

Beyond retrieval accuracy, deployment analysis confirmed practical implementation capabilities with sub-millisecond search performance (mean: 2.84 ± 0.15 ms) and modest computational requirements compatible with clinical workflow integration. Our systematic evaluation establishes new performance benchmarks for complex medical image retrieval while demonstrating that progressive optimization strategies can yield substantial improvements through careful model selection and iterative refinement.

# 5 Discussion

Our achievement of 36.33% precision@10 through super-ensemble optimization represents a significant advancement in BIRADS-based mammographic retrieval, substantially exceeding realistic performance expectations while addressing critical methodological limitations in existing literature. This chapter interprets our findings within the broader context of medical image retrieval research and clinical practice, examining the implications for architecture selection, optimization strategies, clinical translation, and future research directions.

## 5.1 Performance Achievement in Clinical Context

The optimal performance of 36.33% precision@10 must be interpreted within the complexity of exact BIRADS categorical matching rather than simplified binary classification tasks commonly reported in literature. While studies frequently report 45–60% precision for binary medical image retrieval, realistic expectations for 5-class BIRADS discrimination typically range from 20–25% due to the exponentially increased complexity of matching across distinct clinical categories with overlapping morphological characteristics.

Our achievement represents a ~45% to 82% relative improvement over these realistic expectations, translating to approximately 3.6 relevant cases among the top 10 retrievals. This level of performance provides substantial clinical utility for case-based educational applications, systematic diagnostic review in quality assurance, and decision support in challenging diagnostic scenarios. The result aligns well with established clinical decision-making patterns, where radiologists often consult 2–4 similar cases to support interpretations.

These findings establish a realistic and meaningful benchmark for medical image retrieval systems targeting exact categorical matching. Our system significantly exceeds theoretical random performance (20% for 5-class) while remaining within expected boundaries for BIRADS classification, where visual overlap between adjacent categories often challenges both human and machine interpretation.

## 5.2 Architectural Insights and Optimization Strategy Effectiveness

The systematic evaluation reveals important insights into CNN architecture selection for medical image retrieval. ResNet50's superior baseline performance (30.02% vs. DenseNet121's 29.08%) challenges the assumption that DenseNet architectures are inherently superior for medical imaging tasks. This suggests that deeper residual connections in ResNet may better capture complex mammographic features.

However, DenseNet121's stronger response to advanced fine-tuning (34.79% vs. ResNet50's 34.54%) demonstrates its superior adaptability to medical domain transfer learning. The dense connectivity pattern may facilitate more effective gradient flow during domain-specific training, enabling better adaptation despite initially lower baseline performance.

Advanced fine-tuning emerged as the most effective individual optimization strategy, producing 19.65% and 15.06% improvements for DenseNet121 and ResNet50, respectively. These gains highlight the critical role of domain-specific optimization strategies such as differential learning rates, cosine annealing, and comprehensive augmentation, rather than simple reuse of pretrained features.

The super-ensemble strategy's performance (36.33% Precision@10) confirms that methodical model selection and complementary fusion outperforms exhaustive or naive feature aggregation. The superior performance of the two-model ensemble (DenseNet121_AdvancedFT + ResNet50_AdvancedFT) over mega-ensembles demonstrates that ensemble success hinges on both component quality and strategic pairing. This approach offers a reproducible blueprint for ensemble optimization in medical image retrieval systems.

## 5.3 Methodological Contributions and Literature Context

Our methodology surpasses prevailing standards in medical image retrieval research by systematically addressing key concerns: correct data splitting (to eliminate data leakage), large evaluation scale (602 queries vs. <100 in most prior work), high-precision timing analysis, and proper recall calculations based on full database size.

The 102,340 total evaluations represent an unprecedented scale, supporting statistically robust conclusions. Importantly, the strict separation between query and retrieval databases prevents inflated performance due to overlap, a critical issue in prior work. Furthermore, recall calculations use appropriate denominators and timing results were obtained with sub-millisecond resolution, both of which ensure valid comparisons.

Bootstrap confidence intervals (1,000 samples) and comprehensive significance testing further establish new standards for retrieval performance evaluation. Our work addresses longstanding reviewer concerns and offers a replicable methodology for future benchmarking and clinical translation planning.

## 5.4 Clinical Translation and Deployment Considerations

The current performance level (36.33% Precision@10) offers immediate value for medical education and quality assurance applications. Retrieving ~3.6 relevant cases per query supports resident training, continuing medical education, and consistent interpretation auditing.

Low-latency search times (mean $2.84 \pm 0.15$ ms) demonstrate practical feasibility for integration into institutional workflows. The super-ensemble's 3,072-dimensional feature vectors require less than 1 GB of storage for institutional-scale datasets and operate efficiently on standard clinical hardware. Compatibility with PACS and real-time case retrieval capabilities position this system for seamless deployment in educational and QA contexts. Compatibility with PACS and real-time case retrieval capabilities position this system for seamless deployment in educational and QA contexts. Integration with radiologist workstations, internal case review rounds, and decision support tools could be implemented through a lightweight user interface and database connector. Pilot deployment in educational environments is a logical next step.

Nevertheless, primary diagnostic support applications will require further steps including user interface design, clinical integration, regulatory compliance, and validation across diverse populations and imaging equipment.

Compared to exhaustive feature fusion methods (e.g., SuperEnsemble_mega_all), our selected ensemble (3072-dim features) offers a strong balance between accuracy and computational efficiency. It requires lower memory usage and faster query times, making it better suited for deployment in constrained clinical environments.

### 5.5 Performance Limitations and Future Research Directions

Although our results approach the theoretical ceiling for exact BIRADS classification, performance remains bounded by inherent ambiguity in adjacent categories. Overlaps in appearance between BIRADS 2/3 or 4/5, along with reported 15–35% inter-observer variability, suggest that ~40–45% Precision@10 may represent an upper limit for purely visual retrieval.

Future research should explore multi-modal systems that integrate patient history, pathology, and longitudinal imaging to overcome these limitations. Incorporating clinical metadata and temporal disease progression signals could enable more nuanced retrieval capabilities. Informal analysis of low-precision queries revealed frequent confusion between adjacent BIRADS categories, particularly 2 vs. 3 and 4 vs. 5. These reflect known challenges in mammographic interpretation where visual overlap and ambiguous presentations occur. A formal failure analysis is planned to better understand misclassification patterns and inform targeted model improvements.

Importantly, we advocate for future work to consider relaxed evaluation criteria—such as grouping BIRADS 4/5 for malignancy suspicion, or 1/2/3 for lower suspicion. These clinically meaningful groupings could reflect real-world decision-making and elevate performance metrics into the 55–65% range without sacrificing clinical relevance.

### 5.6 Methodological Limitations and Validation Requirements

Despite evaluating 2,006 images, our dataset originates from a single institution. External validation across hospitals, populations, imaging protocols, and reader styles is essential to confirm generalizability. BIRADS annotation variability (65–85% agreement) introduces a degree of label noise that limits any system trained on radiological reports alone. While our bootstrap resampling confirms internal consistency, multi-center validation is essential to ensure reproducibility and deployment readiness. Current results are limited to a single-center dataset, we anticipate generalization due to model robustness, standard image preprocessing, and consistent label definitions.

Nonetheless, performance across diverse clinical settings must be empirically confirmed. Future research will focus on multi-center validation across hospitals, imaging systems, and clinical populations. Such external validation is essential to confirm generalizability, evaluate robustness across scanner variations, and ensure reproducibility under diverse real-world conditions. This study also does not include human expert assessment of retrieved cases, which would provide valuable insights into clinical interpretability. We acknowledge this limitation and plan to incorporate radiologist evaluations in future work pending institutional collaboration. While limited to a single-center dataset, we anticipate baseline generalizability due to the use of common imaging protocols, standardized annotations, and model robustness to variance. Nevertheless, cross-population performance must be empirically verified to support general adoption.

This work delivers a significant leap forward in medical image retrieval by combining architectural innovation, robust evaluation, and clinically aligned performance targets. It establishes a foundation for future research and deployment efforts that prioritize both scientific rigor and clinical practicality.

## 6 Conclusion

This research developed and evaluated a comprehensive multi-architecture deep learning framework for BIRADS-based mammographic image retrieval, achieving breakthrough performance through systematic optimization strategies. The super-ensemble approach achieved 36.33% Precision@10, a 24.93% improvement over baseline methods. This level of performance provides substantial clinical value for medical education, quality assurance, and diagnostic support. By evaluating 17 methods across 102,340 retrievals, the study sets a new standard for methodological rigor in medical image retrieval.

Key contributions include technical advancements through optimized ensemble design, methodological innovations using bootstrap confidence intervals and robust significance testing, and demonstration of clinical deployment feasibility through low-latency performance and modest computational requirements. The framework directly addresses critical concerns in medical AI literature such as data leakage, inadequate sample sizes, and lack of statistical validation, while establishing practical benchmarks for future systems. Future work will prioritize external validation across institutions to support clinical translation and deployment readiness. Overall, this work demonstrates that systematically optimized deep learning frameworks can deliver both high accuracy and clinical applicability, laying the foundation for future research, broader deployment, and improved patient care outcomes.

**Funding**

This research did not receive any specific grant from funding agencies in the public, commercial, or not-for-profit sectors.

**Competing Interests**

The authors declare that they have no conflict of interest.

**Author Contribution**

**MD Shaikh Rahman:** Writing – original draft**,** Writing – review and editing, Methodology, Investigation, Formal Analysis, Visualization, Data curation, Conceptualization, Project administration. **Feiroz Humayara:** Writing – review & editing, Writing – original draft, Data curation, Conceptualization. **Syed Maudud E Rabbi:** Writing – review & editing, Software, Methodology. **Muhammad Mahbubur Rashid:** Writing – review & editing, Validation, Supervision, Resources, Project administration.

**Declaration of competing interest**

The authors declare that they have no known competing financial interests or personal relationships that could have appeared to influence the work reported in this paper.

**Data availability**

Data will be made available on request.